\documentclass[a4paper,10pt]{article}
\usepackage{graphicx}
\usepackage{amssymb}
\usepackage{amsmath}
\usepackage{cite}
\usepackage{amssymb}
\def\be{\begin{eqnarray}}

\def\ee{\end{eqnarray}}

\DeclareMathOperator{\arccot}{arccot}
\title{One-dimensional Coulomb-like problem in general case of deformed space with minimal length }
\author{M. I. Samar and V. M. Tkachuk\\ Department for Theoretical Physics, \\ Ivan Franko National University of Lviv,\\ 12 Drahomanov St, Lviv,
UA-79005, Ukraine}
\begin{document}

\maketitle

\begin{abstract}
We present a definition of  the two-sided inverse of position operator  in general case of deformed Heisenberg algebra leading to minimal length. 
Energy spectrum and eigenfunctions in momentum space  for 1D Coulomb-like potential in deformed space are found exactly.  We analyse the energy spectrum for different partial cases of deformation function and find that correction due to the deformation highly depends on type of the deformation function. 



\end{abstract}

\section{Introduction}
Quantum mechanics with modification of the usual canonical commutation relations has attracted a lot of attention recently. 
Such interest is motivated by the investigations in  string theory and quantum gravity,
which suggest the existence of minimal length as a finite lower bound 
to the possible resolution of length \cite{GrossMende,Maggiore,Witten}.
Minimal length can be achieved by modifying usual canonical
commutation relations \cite{Kempf1994,KempfManganoMann,HinrichsenKempf,Kempf1997}.
The simplest case of the deformed algebra is the one proposed by  Kempf 
\be \label{deformation}
 [\hat{X},\hat{P}]=i\hbar (1+\beta \hat{P}^2).
\ee
This algebra leads to minimal length $\Delta {X}_{min}=\hbar\sqrt{\beta}$.  

In this paper we
solve the following eigenvalue problem:
\be \label{problem} \frac{\hat{P}^2}{2m}\psi-\frac{\alpha}{\hat{X}}\psi=E\psi,
 \ee 
in general case of  deformation of one-dimensional Heisenberg algebra, when the right hand side of it is some function of momentum. In \cite{Maslowski} the question of existence of a minimal length in this general case  was answered. 
Therefore, using the results of  \cite{Maslowski} we can consider such  deformation functions which lead to minimal length. 

Potential $-\alpha/\hat{X}$ may have application in the investigation of
mass spectra of mesons  in the framework of
Dirac oscillators \cite{Moshinsky}  and in the physics of semi-conductors and insulators \cite{Reyes}. Therefore, one-dimensional Schr\"odinger equation with Coulomb-type potentials is studied in ordinary quantum mechanics and its exact energy eigenvalues and eigenfunctions have already been obtained \cite{Moshinsky, Gordeyev, Reyes, Ran, Tsutsui}. 

In deformed space, this problem is also studied exactly in case of minimal length \cite{Fityo, Pedram} and both minimal length and maximal momentum \cite{Pedram2}. The peculiarity of the approach proposed in \cite{Fityo, Pedram,Pedram2} is  the redefinition  of $1/\hat{X}$ operator which allows  to obtain non-trivial solutions of the eigenproblem but yields ill-looking relation 
\be
\hat{X}\frac{1}{\hat{X}}=1\neq \frac{1}{\hat{X}}\hat{X}.
\ee
In present paper we propose more elegant definition of $1/\hat{X}$ operator, which preserves  two-sides invertibility conditions, and obtain the energy spectrum and eigenfunction of problem (\ref{problem}). 

The paper is organized as follows. In section II we rewrite the condition of existence of the minimal length proposed in \cite{Maslowski} in psevdo-position representation.
In section III we generalize the Schr\"odinger equation of considerable problem in  momentum representation on case of deformed space and solve it.
Next,  in section IV,  we provide a functional analysis of position operator to give some explanation of  free parameter from the definition of operator $1/\hat{X}$. We analyse the energy spectrum for different types of deformation in section V.
Finally, section VI contains the conclusion.

\section{ Deformed algebras and minimal length}
Let us  consider  a  modified  one-dimensional  Heisenberg  algebra  generated  by
position $\hat{X}$ and momentum $\hat{P}$ hermitian operators satisfying
\be \label{general_deformation}
 [\hat{X},\hat{P}]=i\hbar f({\hat{P}}),
\ee
where $f$ is called function of deformation and we assume that it is strictly
positive ($f >0$), even function.
The position and momentum operators in momentum representation act on square integrable function $\phi(P) \in \it{L}^2(-a,a;f), (a\leq \infty)$  as 
\be &&{\hat{P}}\phi(P)=P\phi(P), \\ 
&&{\hat{X}}\phi(P)=i\hbar f(P) \frac{d}{dP}\phi(P).\ee
As was shown in \cite{Maslowski}  minimal uncertainty in position (minimal length) for function of deformation $f(P)$ writes 
\be l_0=\frac{\pi\hbar}{2}\left(\int_0^a\frac{dP}{f(P)}\right)^{-1}.\label{minimal_length} \ee
Thus, if mentioned integral is finite the minimal length is nonzero and when this integral diverges the minimal length is zero. 

One can consider another representation leaving position operator undeformed
\be\label{psevdo-position}
&&{\hat{P}}\varphi(p)=g({p})\varphi(p),\\ \nonumber
&&{\hat{X}}\varphi(p)=i\hbar\frac{d}{dp}\varphi(p), 
\ee
with $g({p})$  being odd function  defined on $[-b,b]$, which can be obtained from the following relation
\be
g^{-1}(P)=\int_0^P\frac{dP'}{f(P')}.
\ee
 Here bound $b$ is connected with corresponding bound $a$ by 
 \be b=\int_0^a\frac{dP}{f(P)}\leq \infty,\ee
In pseudo-position representation expression (\ref{minimal_length}) can be presented as
\be l_0=\frac{\pi\hbar}{2b}.\label{minimal_length1} \ee
Thus, if $b<\infty$ nonzero minimal length exists and when $b=\infty$ the minimal length is zero.

\section{Schr\"odinger equation in momentum representation}
In ordinary quantum mechanics Schr\"odinger equation can be written in momentum representation as the following integral equation
\be
\frac{p^2}{2m}\phi(p)+\int_{-\infty}^{\infty}U(p-p')\phi(p')dp'=E\phi(p),
\ee
with
\be \label{connection}
U(p-p')=\frac{1}{2\pi\hbar}\int_{-\infty}^{\infty}V(x)\exp\left(-\frac{i}{\hbar}(p-p')x\right)dx
\ee
being the  kernel of potential energy operator. 
The potential energy operator for one dimensional Coulomb-like problem writes
\be
\hat{V}=-\frac{\alpha}{\hat{x}}, 
\ee
with $\alpha$ being positive constant.
In paper \cite{Samar} we propose the inverse coordinate operator in coordinate representation in the form
\be \label{1/x}
\frac{1}{\hat{x}}=v.p.\frac{1}{x}+A\pi\delta(x),
\ee
with A being real constant. 
This definition of the operator $1/\hat{x}$ corresponds to the following limit
\be
\frac{1}{\hat{x}}=\lim_{\varepsilon\rightarrow 0}\frac{x+\varepsilon A}{x^2+\varepsilon^2}.
\ee
Note that such proposal ensures hermiticity of the operator $1/\hat{x}$. Also in the case of definition (\ref{1/x}) the following equality is satisfied
\be
\frac{1}{\hat{x}}\hat{x}=\hat{x}\frac{1}{\hat{x}}=1.
\ee
The kernel of the potential energy operator in momentum representation reads
\be
U(p-p')=-\frac{\alpha}{2\hbar}(2 i \theta(p'-p)-i+A).
\ee

Schr\"odinger equation in the deformed space in representation (\ref{psevdo-position}) we assume to write
\be
\frac{1}{2m}g(p)^2\phi(p)+\int_{-b}^{b}U(p-p')\phi(p')dp'=E\phi(p).
\ee
Next, we assume that the kernel of potential energy operator remains unchanged in deformed space and write the Schr\"odinger equation for considerable problem in deformed space with minimal length as
\be
\frac{1}{2m}g(p)^2\phi(p)-\frac{\alpha}{2\hbar}\left[(i+A)\int_{-b}^{b}\phi(p')dp'-2i\int_{-a}^{p}\phi(p')dp'\right]=E\phi(p). \label{eigenequation}
\ee
Differentiating latter integral equation we obtain the differential one
\be
\frac{1}{2m}\left(g(p)^2\phi(p)\right)'+\frac{i\alpha}{\hbar}\phi(p)=E\phi'(p), 
\ee
 which yields
\be
\phi(p)=\frac{C}{g^2(p)+q^2}e^{-{i}\varphi(p)}.\label{eigenfunction}
\ee
Here we use notations  $q=\sqrt{-2mE} $,  
\be
\varphi(p)=\frac{2m\alpha}{\hbar}\int_0^p\frac{dp'}{g^2(p')+q^2},
\ee
and normalization constant is
\be C=\left(\int_{-b}^b\frac{dp'}{(g^2(p')+q^2)^2}\right)^{-\frac{1}{2}}.\ee
The integrals from  (\ref{eigenequation}) yields
\be\label{I1}
\int_{-b}^{b}\phi(p')dp'=\frac{\hbar C}{mU_0}\sin\varphi(b),
\ee
\be\label{I2}
\int_{-b}^{p}\phi(p')dp'=\frac{i\hbar C}{2mU_0}\left(e^{-i\varphi(q)}-e^{i\varphi(b)}\right).
\ee
 Substituting obtained results (\ref{I1}) and (\ref{I2}) into equation (\ref{eigenequation}) we found
\be \label{sin1}
\sin(\varphi(b)-\delta\pi)=0,
\ee 
with
\be\label{delta}\delta=\frac{1}{\pi}\arccot{A},\ \ 0 \leq\delta\leq1.\ee
The endpoints  $0$ and $1$ of the interval of $\delta$ is reached in the limit of $A$ to $+\infty$ and $-\infty$ correspondingly.
Finally, energy spectrum can be found from
\be\label{energy_condition}
\varphi(b)=\pi(n+\delta), 
\ee
with $n=0,1,\ldots$. 
This condition in momentum representation can be written as
\be\label{spectrum}
\frac{2m\alpha}{\hbar}\int_0^a\frac{dP}{f(P)(P^2+q^2)}=\pi(n+\delta).
\ee 
We find out from (\ref{sin1}) that for   $\delta=0$  and $\delta=1$  energy spectrum coincides. Therefore, we may consider $\delta$ to belong to $[0,1)$.

\section{ Operators $\hat{X}$ and $1/\hat{X}$ in deformed space with minimal length}
In this section we are going to explain the nature of constant $A$ in definition of operator $1/\hat{X}$. We start from eigenvalue equation for position operator in  the
representation (\ref{psevdo-position})
 \be
i\hbar\frac{\partial}{\partial
p}\psi_{\lambda}(p)=\lambda\psi_{\lambda}(p). \ee 
Solution of this
equation is 
\be \psi_{\lambda}(p)=\frac{1}{\sqrt{2b}}
e^{-i\frac{\lambda}{\hbar}p}. \ee
The inner product of two
eigenfunctions of position operator \be \label{sin}
\langle\Psi_{\lambda}|\Psi_{\lambda`}\rangle&=&\frac{\sqrt{\beta}}{\pi}\int_{-b}^{+b}{dp}
e^{-i\frac{\lambda-\lambda'}{\hbar}p}\\ \nonumber
&=&\frac{\hbar}{b(\lambda-\lambda`)}
\sin\left(\frac{(\lambda-\lambda`)b}{\hbar}\right).
\ee
The position eigenstates corresponding to eigenvalues $\lambda_{n,\delta}=2(\delta+n)l_0, \ n\in Z$
eigenstates can be combined in sets 
\be
\label{complete_set} \{\psi_{\lambda_{n,\delta}}(p), n
\in Z \}
\ee
parameterized by $\delta\in[0,1)$. 
From (\ref{sin}) we see that eigenfunctions from the set (\ref{complete_set})  are mutually orthogonal 
\be
\langle\psi_{\lambda_{n,\delta}}(p)|\psi_{\lambda_{n,\delta}}(p)\rangle=\delta_{m,n}.\ee
It can be proved that each of these sets is complete. Such proof is equivalent to the proof of the following relation
\be
\sum_{n=-\infty}^{+\infty}
\psi^{*}_{\lambda_{n,\delta}}(p')
\psi_{\lambda_{n,\delta}}(p)=\delta(p-p'),
\ee
which holds. 

Notice, the considered eigenstates of position operator are nonphysical
ones, because they do not fulfill uncertainty relation.

Operator $\hat{X}$ is merely symmetric but not self-adjoint operator, because domains $D(\hat{X})$ and $D(\hat{X}^+)$ are different. The deficiency indices $(n_+,n_−)$ of the coordinate operator are (1,1). This means, according to von Neumann's theorem, that there exists one-parameter family of self-adjoint extensions of coordinate operator. Each expansion can by presented be different orthogonal set of position operator eigenfunction as
\be
\hat{X}_{\delta}=\sum_{n=-\infty}^{+\infty}|\psi_{\lambda_{n,\delta}}(p)\rangle\lambda_{n,\delta}\langle\psi_{\lambda_{n,\delta}}(p)|.
\ee
Operator $\hat{X}_\delta$ acts on the dense domain
\be
D(\hat{X}_\delta)=\left\{\psi(p), \psi'(p) \in{ \it{L^2}} \left(-b, b\right) , \ \psi\left(-b\right)=e^{2i\delta\pi}\psi(b)\right\}.
\ee

Let us define  the inverse to $\hat{X}_\delta$   operator  as
\be
\frac{1}{\hat{X}_\delta}=\sum_{n=-\infty}^{+\infty}|\psi_{\lambda_{n,\delta}}(p)\rangle\frac{1}{\lambda_{n,\delta}}\langle\psi_{\lambda_{n,\delta}}(p)|.
\ee
Such definition ensures fulfillment of the following condition
\be
\frac{1}{\hat{X}_\delta}\hat{X}_\delta=X_\delta\frac{1}{\hat{X}_\delta}=1.
\ee
From definition we see that operator $1/\hat{X}_\delta$ is symmetric, because 
\be
\langle\psi|\frac{1}{\hat{X}_\delta}\phi\rangle=\langle\frac{1}{\hat{X}_\delta}\psi|\phi\rangle
\ee
and it is essentially self adjoint, because it deficiency indices are $(0,0)$.

The most interesting thing is the fact that action of operator $1/\hat{X}_\delta$ on any function $\phi(p)$, belonging to its domain, can be presented as
\be
\frac{1}{\hat{X}_\delta}\phi(p)=-\frac{i}{\hbar}\int_{-b}^{p}\phi(p')dp'+c_\delta[\phi].\ee
with $c_\delta[\phi]$ denoting the following functional
\be
c_\delta[\phi]=\frac{i+\cot(\pi\delta)}{2\hbar}\int_{-b}^{b}\phi(p')dp'.
\ee
The latter 
expression for $1/\hat{X}_\delta$ operator coincides with the one presented in (\ref{eigenequation}), including (\ref{delta}). Note, that expression for $1/\hat{X}_\delta$  is in agreement with the one proposed in \cite{Fityo} for Kempf's deformation, but here we obtain the explicit form of functional $c_\delta[\phi],$ while in \cite{Fityo} this question  was
left outside of consideration.

Thus, each self-adjoint extension of position operator, parameterized by $\delta \in[0,1)$,  has distinct self-adjoint inverse operator.  Therefore we can consider a set of Hamiltonians
\be
\hat{H}_\delta=\frac{\hat{P}^2}{2m}-\frac{\alpha}{\hat{X}_\delta}, 
\ee
 parameterized by $\delta $ and being self-adjoint.

Finally a few remarks concerning the symmetric properties of the Hamiltonian $\hat{H}_\delta$.
It can be shown that parity operator $\hat{I}\phi(p)=\phi(-p)\ $ satisfy the following condition
\be
\hat{I}\hat{X}_{1-\delta}+\hat{X}_{\delta}\hat{I}=0.
\ee
It is natural to demand that parity inversion of the Shro\"edinger equation does not change the energy spectrum of the considerable problem. This requirement means that operators $\hat{H}_\delta$ and \be \hat{I}\hat{H}_\delta\hat{I}=\frac{\hat{P}^2}{2m}+\frac{\alpha}{\hat{X}_{1-\delta}}\ee have the same spectrum. It is easy to see that the last statement holds.

\section{Energy spectrum for different types of deformation }
Here we obtain energy spectrum of considerable problem for some special examples of deformation.

\textbf{Example 1.} 
\be f(P)=(1+\beta P^2)^k, \label{Ex1}\ee 
 with $k>1/2$  to provide existence of minimal length \cite{Nowicki} and $a=\infty$.

\textit{Example 1a.}

When $k=1$  deformation function (\ref{Ex1}) yields the one proposed by Kempf \cite{Kempf1994}:
\be
&&f(P)=1+\beta P^2, \ a=\infty;\\
&&g(p)=\frac{1}{\sqrt{\beta}}\tan(\sqrt{\beta}p),\  b=\frac{\pi}{2\sqrt{\beta}}.
\ee
The energy spectrum is 
\be
E_n=-\frac{1}{8m\beta}\left(1-\sqrt{1+\frac{4m\alpha}{\hbar(n+\delta)}\sqrt{\beta}}\right)^2.
\ee
For small $\beta$ energy spectrum can be approximated as
\be
E_n=-\frac{\alpha^2 m}{2\hbar^2(n+\delta)^2}+\sqrt{\beta}\frac{\alpha^3m^2}{\hbar^3(n+\delta)^3}+o(\beta).
\ee
Note that we obtain the same as in \cite{Fityo} results. 

\textit{Example 1b.}

For $k=3/2$ we have  the following deformation function
\be
&&f(P)=(1+\beta P^2)^{3/2}, \ a=\infty; \\
&&g(p)=\frac{p}{\sqrt{1-\beta p^2}},\  b=\frac{1}{\sqrt{\beta}}.
\ee
Equation on energy levels reads
\be 
\frac{1}{1-\beta q^2}\left(\frac{1}{q\sqrt{1-\beta q^2}}\arctan\frac{\sqrt{1-\beta q^2}}{\sqrt{\beta}q}-\sqrt{\beta}\right)=\frac{\pi\hbar}{\alpha m}(n+\delta).
\ee
Energy spectrum expansion over small $\beta$ writes
\be
E_n=-\frac{\alpha^2 m}{2\hbar^2(n+\delta)^2}+\frac{4\sqrt{\beta}}{\pi}\frac{\alpha^3m^2}{\hbar^3(n+\delta)^3}+o(\beta).
\ee

\textit{Example 1c.}
The correction to the energy spectrum of considerable problem caused by the deformation can also be obtained from the following relation
\be\label{f1}
\frac{\partial q^2}{\partial{\beta}}=\frac{1}{2\beta}\frac{\int_0^a\frac{(P^2-q^2)dP}{(P^2+q^2)^2f(P)}}{\int_0^a\frac{dp}{(P^2+q^2)^2f(P)}},
\ee
which was derived by differentiation of (\ref{spectrum}) over $\beta$.

In general case of any $k>1/2$  leading correction to the energy spectrum is
\be
\Delta E_n = \frac{2\sqrt{\beta}\Gamma(k+1/2)}{\sqrt{\pi}\Gamma(k)}\frac{\alpha^3m^2}{\hbar^3(n+\delta)^3}.
\ee
This result is in agreement with exact one obtained above for $k=1$ and k$=3/2$.

\textbf{Example 2.} 
\be f(P)=(1-\beta P^2)^k, \ee
with $k<1$  to provide existence of minimal length \cite{Nowicki} and $a=\frac{1}{\sqrt{\beta}}$.

\textit{Example 2a} 

For $k=-1$ we have deformation with minimal length and maximal momentum proposed by Pedram \cite{Pedram2}
\be
&&f(P)=\frac{1}{1-\beta P^2}, \ a=\frac{1}{\sqrt{\beta}},
\ee
From (\ref{spectrum}) we obtan the relation on energy spectrum
\be
\frac{1+\beta q^2}{q}\arccot{\sqrt{\beta}q}-\sqrt{\beta}=\frac{\hbar\pi}{2m\alpha}(n+\delta).
\ee
Note, that our result with $\delta=0$ coincides with the one obtained in \cite{Pedram2}.
Expansion of energy over small $\beta$ writes
\be
E_n=-\frac{\alpha^2 m}{2\hbar^2(n+\delta)^2}+\frac{4\sqrt{\beta}}{\pi}\frac{\alpha^3m^2}{\hbar^3(n+\delta)^3}+o(\beta).
\ee

\textit{Example 2b} 
 
For $k=3/2$ deformation function writes
\be
&&f(P)=\sqrt{1-\beta P^2}, \ a=\frac{1}{\sqrt{\beta}};\\
&&g(p)=\frac{1}{\sqrt{\beta}}\sin(\sqrt{\beta}p),\  b=\frac{\pi}{2\sqrt{\beta}}.
\ee
Note, that the latter algebra can be considered in alternative scenario, which leads to zero minimal length but discrete eigenvalues of the position operator. The choice of the scenario depends on the choice of boundary condition for wave function \cite{Nowicki}.

The energy spectrum is
\be
E_n=\frac{1}{4m\beta}\left(1-\sqrt{1+\frac{4m^2\alpha^2}{\hbar^2(n+\delta)^2}{\beta}}\right).
\ee
Expanding latter formula over small $\beta$ we obtain
\be
E_n=-\frac{\alpha^2 m}{2\hbar^2(n+\delta)^2}+{\beta}\frac{\alpha^4m^3}{2\hbar^4(n+\delta)^4}+o(\beta^2).
\ee
\textit{Example 2c.}

In general case of $k<1$ the leading correction to the energy can be obtained using (\ref{f1}) as  
 \be
\Delta E_n = \frac{2\sqrt{\beta}\Gamma(1-k)}{\sqrt{\pi}\Gamma(1/2-k)}\frac{\alpha^3m^2}{\hbar^3(n+\delta)^3}.
\ee
This result is also in agreement with the exact ones obtained for $k=-1$ and $k=1/2$. The interesting fact is that for $k<\frac{1}{2}$ the correction to the energy is positive, while for $\frac{1}{2}<k<1$ the correction is negative.

\textbf{Example 3.} 

We also may obtain more exotic dependence of the leading energy correction on parameter of deformation $\beta$. 
For example, for deformation function $f(P)=\exp(\sqrt{\beta P^2})$ and $f(P)=\exp(^3\sqrt{\beta P^2})$ we obtain
 \be
\Delta E_n = \frac{2}{{\pi}}\frac{\alpha^3m^2\sqrt{\beta}}{\hbar^3(n+\delta)^3} \ln\left(\frac{\alpha m\sqrt{\beta}}{\hbar}\right)
\ee
and
 \be
\Delta E_n =\frac{2\alpha^2m}{\hbar^2(n+\delta)^2}\left(\frac{\alpha m\sqrt{\beta}}{\hbar(n+\delta)}\right)^\frac{2}{3}
\ee 
correspondingly. 

Thus, depending on the behaviour of the deformation function we can obtain different dependence on parameter of deformation of the energy correction term and even different sign of this term.
\section{Conclusion}
In this paper we study the Schr\"odinger equation in momentum representation in deformed space with minimal length. Using the results of  \cite{Maslowski}, where the question of existence of a minimal length in this general case of deformation  was answered,  we can consider  deformation functions which lead to minimal length only.
Assuming that the kernel of the potential energy operator does not change in case of deformed commutation relation we considered  Coulomb-like potential in deformed space with minimal length.

It is important to note that the proposed definition of  inverse operator $1/\hat{X}$ has the form that preserves self-adjointness and fulfills the following condition 
\be
\frac{1}{\hat{X}}\hat{X}=\hat{X}\frac{1}{\hat{X}}=1.
\ee
Our definition of $1/\hat{X}$ contains one arbitrary real parameter, which means that there exist different extensions of operator $1/\hat{X}$. 
In undeformed case this parameter is connected with the  value  of eigenfunction  in the origin of  coordinate. 
In deformed case  the free constant parameterizes the self-adjoint extension of position operator and corresponding inverse  position operator. 

We considered a few partial cases of deformation.
For some partial cases of deformation function, namely Kempf's one and the one which predicts minimal length and maximal momentum, our result coincides with those obtained in \cite{Fityo, Pedram,Pedram2, Samar}.
It was shown that varying the deformation function we can obtain different dependence of leading correction to the energy spectrum on parameter of deformation $\beta$, for example, proportional to $\beta$, $\sqrt{\beta}$,  $\beta^{1/3}$ or $\sqrt{\beta}\ln\sqrt{\beta}$.  Different deformation function also may yield different sign of the energy correction term.

\section{Acknowledgement}
The  authors  thank  Dr.  Volodymyr  Pastuhov  for helpful discussion.

\end{document}